\documentclass{aa}

\usepackage{graphicx}

\begin{document}

\def\i{\,{\sc i}}
\def\ii{\,{\sc ii}}
\def\teff{$T_{\rm eff}$}
\def\lgg{$\log\,{g}$}
\def\vt{$\xi_{\rm t}$}
\def\kms{\,km\,s$^{-1}$}
\def\EqW{$W_\lambda$}
\def\vsini{$v \sin i$} 

 
\titlerunning{Stark broadening on Si\i\ lines}

\title{
On the influence of Stark broadening on Si\i\ lines in
stellar atmospheres}

\author{M. S. Dimitrijevi\'c\inst{1,}\inst2, T. Ryabchikova\inst{3,} 
\inst4,  L. \v
C. Popovi\'c\inst{1,}\inst2,
D. Shulyak\inst5, V. Tsymbal\inst5}

\institute{Astronomical Observatory, Volgina 7, 11160 Belgrade
74, Serbia
\and
 Isaac Newton Institute of Chile, Yugoslavia Branch
\and
Institute of Astronomy, Russian Academy of Science, Pyatnitskaya 48,
119017 Moscow, Russia
\and
Institute for Astronomy, University of Vienna, T\"urkenschanzstrasse 17,
A-1180 Vienna, Austria
\and
Tavrian National University, Yaltinskaya 4, 330000 Simferopol, Crimea,
Ukraine
}
\date{Received 12 November 2002 / Accepted 27 March 2003}

\offprints{M. S. Dimitrijevi\'c (mdimitrijevic@aob.bg.ac.yu)}
\authorrunning{M. S. Dimitrijevi\'c, T. Ryabchikova,  L. \v C. Popovi\'c,
D. Shulyak, V. Tsymal}

\abstract{
We study the influence of Stark broadening and stratification effects on Si\i\,
lines in the rapidly oscillating
(roAp) star 10 Aql, where the Si\i\,  6142.48\ \AA\ and  6155.13 \AA\ lines
are asymmetrical and shifted.
First we have calculated Stark broadening parameters using the semiclassical
perturbation method for three Si\i\,
lines: 5950.2\ \AA, 6142.48\ \AA\ and  6155.13 \AA. We revised the synthetic spectrum
calculation code  taking  into account both Stark width and shift for these
lines. From the comparison of our calculations with the observations we found
that Stark broadening + the stratification effect can explain
asymmetry of the Si\i\,  6142.48\ \AA\ and  6155.13 \AA\
lines in the atmospere of roAp star 10 Aql.
   }

\maketitle

\section{Introduction}

The Stark broadening is the most significant pressure broadening
mechanism  for A and B
stars and one has to take into account this effect
in
investigation, analysis and modeling of their atmospheres.
In one of our previous works (Popovi\'c et al. 2001) we have
shown that the Stark effect may change the equivalent width of spectral
lines by 10-45\%, hence neglecting this mechanism,
we may introduce a significant
errors in abundance determinations.
On the other hand, high resolution spectra allow us to
study different broadening effects using line profiles. In the course of the abundance
analyses of peculiar (Ap) stars we noticed that some of the Si\i\ lines are
shifted relative to the laboratory wavelength. Moreover, few strong lines
mainly from the multiplets $3p^3 \ ^3D^0-5f^3D$ and  $3p^3 \ ^3D^0-5f^3G$ have
asymmetrical line profiles, in particular the Si\i\ 6155.13
\AA\ line. We found that this line is slightly shifted and
asymmetrical even in the solar spectrum, while in the hotter, e.g. Ap
stars, the shift and asymmetry are more pronounced.

The aim of this paper is to explain the asymmetry of Si\i\ lines within the
framework of the Stark broadening effect. First we calculated the Stark
broadening parameters for these lines, after that we calculated synthetic
spectra and compared them with the observations,
and finally, we discuss the effects causing shift and
asymmetry in the line profile of the Si 6155.13 and 6142.48 \AA\ lines.

\section{Observations}

For our analysis we used observations of one normal star HD~32115,
two Ap stars HD~122970 and 10 Aql and Solar Flux Atlas by Kurucz et al.
(1984).
High resolution CCD spectra of 10 Aql and HD~122970 are described  by
Ryabchikova et al. (2000).  High resolution CCD spectra
(R$\approx$45000) of
HD~32115 in the wavelength region 4000--9500 \AA\ were obtained with the
coude-echell spectrometer mounted on the 2m 'Zeiss' telescope at Peak
Terskol Observatory, Russia (see Bikmaev et al. 2002 for more details).

More Ap stars show peculiar line profiles of Si\i\, lines but most 
stars have rather strong magnetic fields which distort line profiles through
Zeeman splitting. Rather weak magnetic fields in Ap stars HD~122970 and 10 Aql
allow us to ignore magnetic effects on line shape.

\section{The Stark broadening parameter calculation}

Calculations have been performed within the semiclassical perturbation
formalism,  developed
and discussed in detail in Sahal-Br\'echot (1969ab). This formalism,
as well as the
corresponding computer code, have been
optimized and updated several times (see e.g. Sahal-Br\'echot,
1974; Dimitrijevi\'c and Sahal-
Br\'echot, 1984a, Dimitrijevi\'c, 1996).

Within this formalism, the
full width of a neutral emitter isolated spectral line broadened by
electron impacts can be expressed in
terms of cross sections for elastic and inelastic processes as

\begin{equation}
W_{if} = {2\lambda_{if}^2\over 2\pi c}n_e\int vf(v)dv (\sum_{i'\ne
i}\sigma_{ii'}(v) + \sum_{f'\ne f}\sigma_{ff'}(v) + \sigma_{el})
\end{equation}
\noindent{and the corresponding line shift as}

\begin{equation}
d_{if} = {\lambda_{if}^2\over 2\pi c}n_e\int vf(v)dv\int_{R_3}^{R_D}
2\pi \rho
d\rho \sin 2\phi_p
\end{equation}

\noindent{Here, $\lambda_{if}$ is the wavelength of the line
originating from the transition with the initial atomic energy level
$i$ and the final level $f$, $c$ is the velocity of light, $n_e$ is the
electron density, $f(v)$ is the Maxwellian velocity distribution
function for electrons, $m$ is the electron mass, $k$ is the Boltzman constant, 
$T$ is 
the temperature, and
$\rho $ denotes the impact parameter of the
incoming electron. The inelastic cross section $\sigma_{jj'}(v)$ is 
determined according to Chapter 3
in Sahal-Br\'echot (1969b), and elastic cross section $\sigma_{el}$ according 
to Sahal-Br\'echot (1969a). The cut-offs, included in
order to maintain for the unitarity of the {\bf $S$}-matrix, are
described in Section 1 of Chapter 3 in Sahal-Br\'echot (1969a).}

The formulae for the ion-impact broadening parameters are
analogous to the formulae for electron-impact broadening. We note that the fact 
that the colliding ions would impact in the far wings should be checked, 
even for stellar densities.

\section{Line profile calculations}

Model atmosphere calculations as well as calculations of the absorption
coefficients
were made with the local thermodynamical equilibrium (LTE) approximation.
Model calculations were performed with the  ATLAS9 code written by R.L. Kurucz (1993).
 The next step is the calculation of the outward flux at corresponding
wavelengths points
using the given model. For this  we used the STARSP program written by
V.V. Tsymbal (1996). In its current state this code includes
the possibility of calculating a synthetic spectrum for an atmosphere with
a vertical stratification of chemical elements.

The computational scheme  is as follows. For each line we find the
central opacity as

\begin{equation} \label{eq:a1}
\alpha_\nu=\frac{\pi e^2}{mc} gf_{if}
e^{-\frac{\chi}{kT}}\frac{n}{\rho}(1-e^{-\frac{h\nu}{kT}}),
\end{equation}

where $\alpha_\nu$ is the mass absorption coefficient at frequency $\nu$, $e$
is the electron charge, $g$ is the statistical
weight, $f_{if}$
is the oscillator strength for a
given transition, $\chi$ is the excitation energy,
 $n$ is the number density of a corresponding element in a given
ionization stage multiplied by partition function, $\rho$ is the density and $h$
is the Planck constant. The last factor describes stimulated emission.

The Doppler width is

\begin{equation}
\Delta \nu_D = \frac{\nu}{c}\sqrt{\frac{2kT}{m_A}+\xi^2_t}.
\end{equation}
In this expression $m_A$ is the  mass of the absorber and $\xi_t$ is the
microturbulent velocity.

Next, we compute the total damping parameter

\begin{equation}
\gamma = \gamma_{rad} + \gamma_{Stark} + \gamma_{neutral}.
\end{equation}

Here $\gamma_{rad}$,  $\gamma_{Stark}$ and $\gamma_{neutral}$ are the
radiative, Stark
and damping parameters due to neutral atom collisions respectively. 
The values of $\gamma_{rad}$,
$\gamma_{neutral}$, excitation energy $\chi$ and oscillator strength $gf$ were taken
from the Vienna Atomic Line Database (VALD) (Kupka et al. 1999).
In the case of neutral atom  broadening we
assumed that perturbing particles are  atoms of neutral hydrogen and helium
only. This assumption is applicable to almost all types of stars due to
high hydrogen and helium cosmic abundances.
 Usually this damping process is called  Van der Waals broadening.
The best theory for atomic hydrogen collisions which includes not only the
Van der Waals potential is given in the
papers by Anstee \& O'Mara (1991) and Barklem \& O'Mara (1998). Barklem, Piskunov \& O'Mara 
(2000) provided $\log\gamma_{neutral}$ 
calculations for about 5000 lines of the neutral atoms and the first ions of many
chemical elements. These damping parameters are included in VALD database
 per one perturbing particle and for a temperature 
of 10000 K. Unfortunately, 
no calculations exist for Si\i\, lines. We shall discuss the competition between 
the broadenings caused by the Stark effect and neutral hydrogen collisions in the 
atmospheres of our stars in Section 5.2.

In order to include Stark broadening effects we added the approximate
formulas (see Eqs. (14) and (15) in Sec. 5.1) in
the code.
The Stark width and shift are

\begin{equation}
\gamma_{Stark} =  \gamma_{Stark}^{(e)}n_e + \gamma_{Stark}^{(p)}n_p +
\gamma_{Stark}^{(HeII)}n_{HeII},
\end{equation}

\begin{equation}
d_{Stark} =  d_{Stark}^{(e)}n_e + d_{Stark}^{(p)}n_p +
d_{Stark}^{(HeII)}n_{HeII},
\end{equation}
where $n_e$, $n_p$ and $n_{HeII}$  are the corresponding densities of 
electrons,
protons and
He\ii\, ions respectively. The  resulting opacity profile 
is given by the Voigt function (Doppler + pressure broadening).

Thus, at each point of a given spectral region (with resolution $0.01$ 
\AA\ for both lines)
we computed line absorption coefficient $\ell_\nu$ as follows

\begin{equation}
\ell_\nu = \alpha_\nu V(u,a),
\end{equation}
where $\alpha_\nu$ is given by Eq. (4) and  $V(u,a)$ is a Voigt function
with parameters

\begin{equation}
u = \frac{\nu-\nu_0+d_{Stark}}{\Delta\nu_D},
\end{equation}
\begin{equation}
a = \frac{\gamma}{4\pi\Delta\nu_D}.
\end{equation}
The Stark shift $d_{Stark}$ and the damping parameter $\gamma$
have been found from (2) and (11), respectively.

The flux is given by the expression

\begin{equation}
H_\nu (\tau_\nu) = -\frac{1}{2} \int_{0}^{\tau_\nu} S_\nu E_2(\tau_\nu-t)dt + 
\frac{1}{2} \int_{\tau_\nu}^{\infty} S_\nu E_2(t-\tau_\nu)dt,
\end{equation}
where $\tau_\nu$ is the optical depth, $S_\nu$ is the 
source function and
$E_n(x)$ is the exponential integral of the order $n$ and argument
$x$. The flux
integral has been evaluated using matrix operators.

\section{Results}

\subsection{Stark broadening data}

The atomic energy levels needed for Stark broadening calculations
were taken from
Martin and Zalubas (1983) and Moore (1971), but LS determination of 
$5f^3D$, $5f^3G$,
$6s^1P^o$ and
7s$^1$P$^o$ terms have been adopted according to
Moore (1971) and therfore, in order to obtain a consistent set of 
data, energy levels from Moore (1971) have been used as the principal  
source.
Oscillator strengths have been
calculated by using the method of Bates and Damgaard (1949) and
tables of
Oertel and Shomo (1968). For
higher levels, the
method described in van Regemorter, Binh Dy and Prud'homme (1979) has
been applied. 

The spectrum of neutral silicon is complex and not known well enough for a good 
calculation of the considered lines. First of all upper and lower energy levels for 
6142.48 \AA\- and 6155.13 \AA\- lines are not known reliably. According to 
Striganov \& Sventickij (1966) they belong to the $3p^3 \ ^3D^0-5f^3D$ and  
$3p^3 \ ^3D^0-5f^3G$ multiplets respectively, while Moore et al. (1966)  
stated that the lower level of the corresponding transitions is $3d^3D^o$. 
It is 
also worth noting that in  
NIST (2002) the lower and upper levels of the corresponding transitions are not 
specified. Consequently, we adopted the identification of Striganov \& 
Sventitskij (1966) as the  only one enabling  the corresponding 
calculations. Moreover, the 
$3s3p^3 \ ^3D_3^o$ level is in fact a mixture of 39\% of $3pnd^3D^o$ and 56\% 
of $3s3p^3 \ ^3D_3^o$ states, and 5\% is unknown. Also the $3pnd^3D^o$  state 
is a mixture of states with 3 $\le n \le$ 12. In calculations, we assumed 
that we have 30\% of $3p3d^3D^o$ state, and our checks show that the difference 
in final results is negligible if we assume that this state is involved up to 20\% 
only. An additional complication was that $g$ levels, which might be very 
important for the perturbation of the considered 5$f$ levels, are unknown. In 
accordance with the decrease of distance between $5s,\- 5p,\- 5d$ and $5f$ 
levels, we estimated that the distance from 5$f^3D$ or 5$f^3G$ term to the 
5$g$ levels should  lie  between 500 and 1500 cm$^{-1}$. We checked 
results 
without 5$g$ levels and with a fictive 5$g$ level at 500, 1000 and 1500 cm$^{-
1}$ from the corresponding 5$f$ terms. In all cases, line widths differed by less than 
1\%, while the shift varies within the limits of 3.5\%. If the 
distance is 500 cm$^{-1}$, the shift value e.g. at 10000 K for 6142.48 \AA\- 
line  differs by about 20\%. If we include a fictive 6$g$ level 500 cm$^{-1}$ 
distant from 5$g$ levels, the difference is negligible.  Hence, in order to 
obtain the needed set of atomic energy levels, we adopted a fictive 5$g$ level 
1000 cm$^{-1}$ distant from the 5$f$ levels and a fictive 6$g$ level 500 cm$^{-1}$ 
distant from the 5$g$ level. Since the average estimated error of the semiclassical 
method is $\pm $30\%, due to additional approximations and uncertainties, we 
estimate the error bars of our results to be $\pm $50\%. 
   
Our results for electron-, proton-, and
ionized helium-impact line widths and shifts for the three considered Si\i\
spectral
lines for a perturber density of 10$^{14}$ cm$^{-3}$
and temperatures T $=$ 2,500 $-$ 50,000 K, are shown in Table 1.
For
perturber densities
lower than those tabulated here, Stark broadening parameters vary
linearly with perturber density. The nonlinear behaviour of Stark broadening
parameters
 at higher densities is the consequence of the influence of Debye
shielding and has been analyzed in detail in Dimitrijevi\'c and
Sahal-Br\'echot (1984b).
 
 \begin{table*}
\begin{center}
      \caption[]{ Stark broadening parameters for Si\i\,
spectral lines.
This table shows electron-, proton-, and ionized
helium-impact
broadening parameters for Si\i\, for a perturber density of  10$^{14}$
cm$^{-3}$ and temperatures from 2,500 up to
50,000 K. The quantity  C (given in \AA\- cm$^{-3}$),
when divided by the corresponding full width at half maximum,
gives an estimate for the maximum perturber density for which
tabulated data may be used. The asterisk identifies cases for which the 
collision volume multiplied by the perturber density (the condition for 
validity of the impact approximation) lies beetwen 0.1 and 0.5. For higher 
densities, 
the isolated line approximation used in the calculations breaks down.
FWHM(\AA) denotes full line width at half maximum in \AA , while SHIFT(\AA)
denotes line shift in \AA.}
\begin{tabular}{|r|r|r|r|r|r|r|r|}
\hline
\multicolumn{2}{|c|}{PERTURBERS ARE:}& \multicolumn{2}{c|}{ELECTRONS}& \multicolumn{2}{c|}{
PROTONS}&\multicolumn{2}{c|}{HELIUM IONS} \\
\hline
TRANSITION    &          T(K)  &  FWHM(A)  &   SHIFT(A) &   FWHM(A)  &   SHIFT(A)  &  FWHM(A)   &
SHIFT(A) \\
\hline
 SiI  $4s - 5p$&  2500.&   0.509E-02 &      0.333E-02 &      0.160E-02 &      
0.701E-03& & \\
   $^1P- ^1D$  &   5000. &  0.581E-02    &   0.409E-02  &     0.172E-02   
&    0.898E-03 & & \\
   5948.55 A  &  10000. &  0.652E-02 &      0.381E-02   &    0.183E-02     &  0.109E-02     &
0.167E-02 &   0.851E-03 \\
 C= 0.36E+19 &  20000.  & 0.740E-02     &  0.328E-02    &   0.194E-02    &   0.127E-02     &
0.175E-02 &  0.101E-02 \\
              & 30000.  & 0.811E-02   &    0.271E-02  &     0.202E-02  &     0.139E-02   &
0.181E-02 &   0.111E-02 \\
              & 50000.  & 0.903E-02     &  0.212E-02    &   0.213E-02     &  0.153E-02  &
0.188E-02 & 0.123E-02 \\
\hline
 SiI  $3p^3-5f$  &  2500. &  0.967E-01 & -0.576E-01 &  0.175E-01 & 
-0.145E-01 & *0.145E-01 &-0.114E-01 \\
      $^3D_3- ^3D_3$  & 5000.  & 0.108   &   -0.624E-01  & 0.196E-01 & 
-0.168E-01  & 0.160E-01 &-0.134E-01 \\
   6142.48 \AA\   & 10000.  & 0.120   &   -0.609E-01  & 0.219E-01 & 
-0.192E-01  & 0.178E-01 & -0.154E-01 \\
   C= 0.52E+16  & 20000.  & 0.132   &   -0.472E-01  & 0.247E-01 & 
-0.219E-01  & 0.198E-01 & -0.176E-01 \\
              & 30000.  & 0.140   &   -0.400E-01  & 0.266E-01 & -0.235E-01  
& 0.211E-01 & -0.189E-01 \\
              & 50000.  & 0.147  &    -0.326E-01  & 0.294E-01 & 
-0.258E-01  & 0.228E-01 & -0.208E-01 \\
 \hline
 SiI  $3p^3-5f$  &  2500.  & 0.905E-01 & -0.622E-01  & 0.184E-01 & 
-0.152E-01  & *0.151E-01 & -0.120E-01 \\
      $^3D_3- ^3G_4$  & 5000.  & 0.102 & -0.707E-01  & 0.204E-01 & 
-0.177E-01  & 0.168E-01 & -0.141E-01 \\
   6155.13 \AA\  &  10000.  & 0.112     & -0.706E-01  & 0.228E-01 & 
-0.203E-01  & 0.187E-01 & -0.162E-01 \\
C= 0.20E+17 & 20000.  & 0.121     & -0.571E-01  & 0.255E-01 
& -0.231E-01  & 0.208E-01 & -0.185E-01 \\
               & 30000.  & 0.129     & -0.482E-01  & 0.273E-01 & 
-0.248E-01  & 0.221E-01 & -0.200E-01 \\
              & 50000.  & 0.137     & -0.392E-01  & 0.298E-01 & 
-0.272E-01  & 0.240E-01 & -0.219E-01  \\
\hline
\end{tabular}
\end{center}
\end{table*}

In Table 1 the Stark widths and shifts for el\-ec\-t-\break ron-, proton- 
and 
He\ii\,
ion-impact broadening are presented as a function of temperature for a 
density
of 10$^{14}$ cm$^{-3}$. After testing the density dependence
of Stark parameters, we have found that the width and shift are linear functions
of  density for perturber densities smaller than 10$^{16}\rm
cm^{-3}$ and can be scaled by the simple formula:

\begin{equation}
(W,d)_N=(W,d)_0({N\over 10^{14}}),
\end{equation}
where $(W,d)_N$ are the width and shift at a perturber 
density $N$
($\rm cm^{-3}$), and
$(W,d)_0$  are width and shift  given in Table 1, respectively.

\begin{figure}
\includegraphics[width=8.8cm]{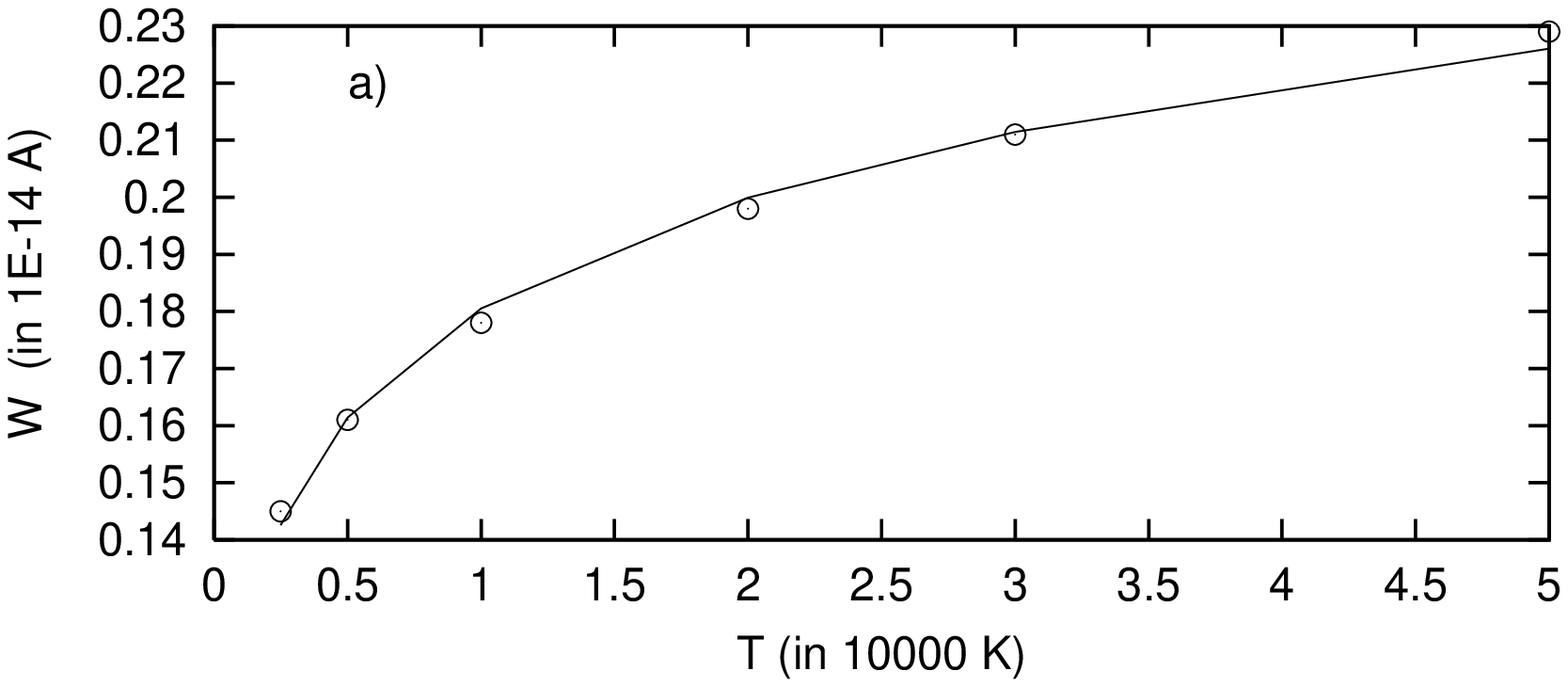}
\includegraphics[width=8.8cm]{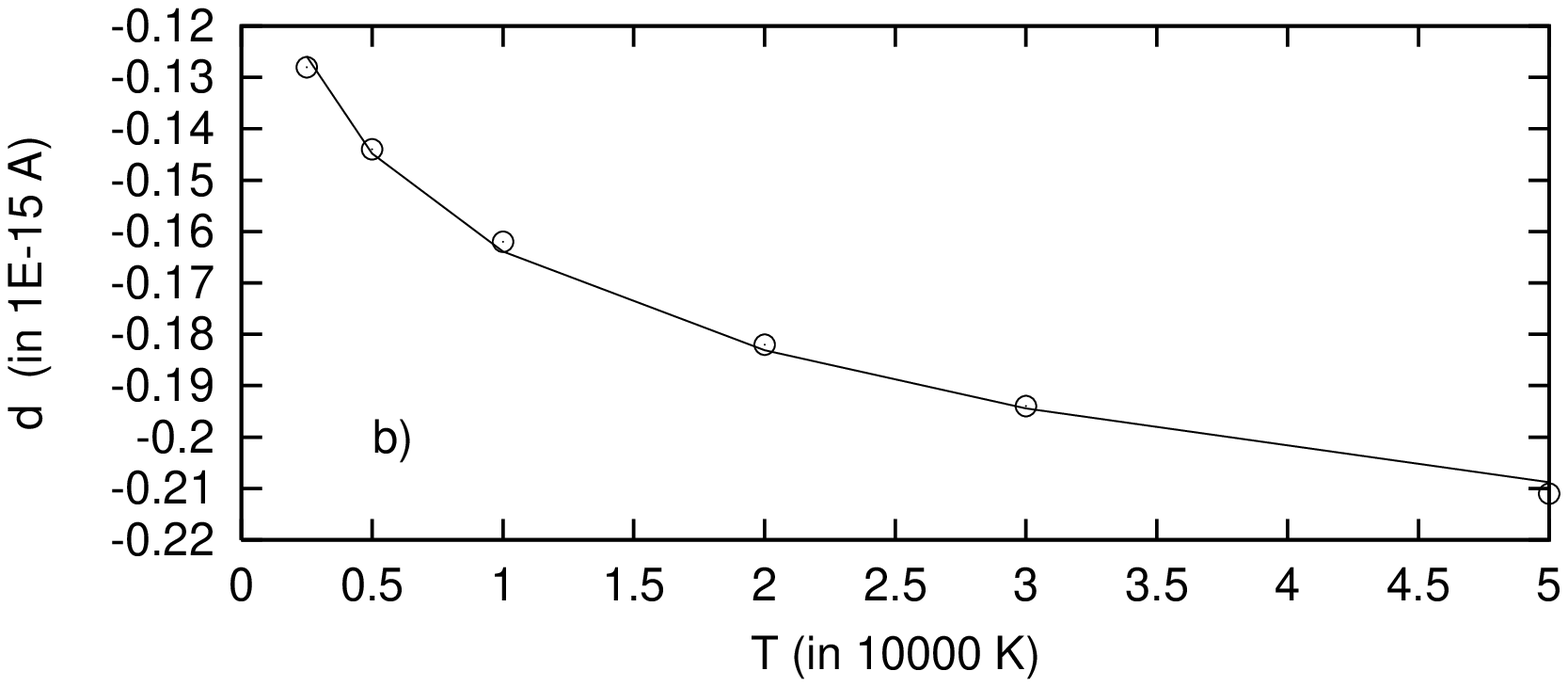}
\caption{The analytic fit of He II-impact broadening data for Si\i\, 6142 
\AA\ line: a)
Stark width, and b) Stark shift. 
The shift and width are given for one perturber per cm$^{3}$}
\end{figure}

\begin{figure}
\includegraphics[width=8.8cm]{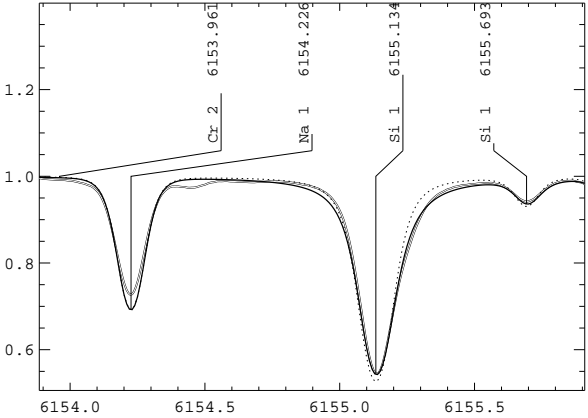}
\caption{A comparison between the observed Si\i\, 6155 \AA\ line profile
in the solar spectrum and synthetic spectra calculated with Stark widths and 
shifts from Table 1 (solid line) and with Stark widths calculated by 
the approximate
formulae and without Stark shift taken into account (dotted line).
X- and Y-coordinates are wavelenghths and surface fluxes (normalized to unity).}
\end{figure}

\begin{figure}
\includegraphics[width=8.8cm]{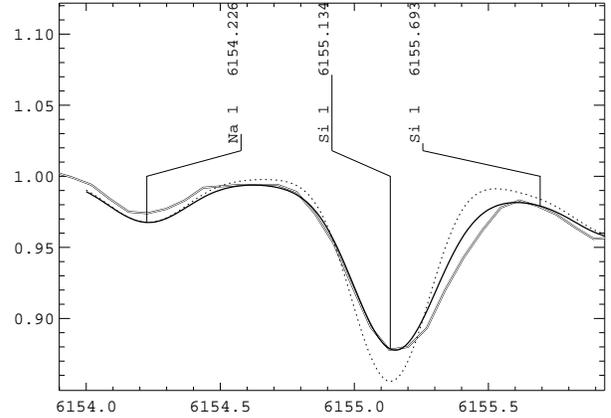}
\caption{The same as in Fig. 2 for normal star HD~32115.}
\end{figure}

\begin{figure}
\includegraphics[width=8.8cm]{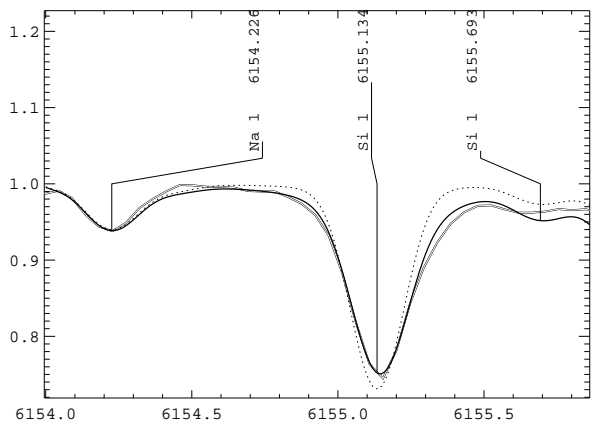}
\caption{The same as in Fig. 2 for Ap star HD~122970.}
\end{figure}

\begin{figure}
\includegraphics[width=8.8cm]{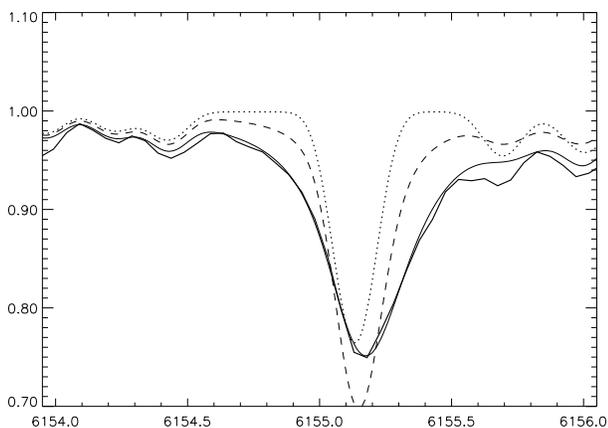}
\caption{A comparison between the observed Si\i\, 6155 \AA\ line profile
in the spectrum of Ap star 10 Aql (thick line) and synthetic spectra 
calculated with Stark widths and shifts from Table 1 and Si abundance 
stratification (thin line), with the same Stark parameters but for homogeneous
Si distribution (dashed line), and with Stark width calculated by approximate
formulae for the same stratification (dotted line). X- and Y-coordinates 
are wavelenghths and surface fluxes (normalized to unity).}
\end{figure}

\begin{figure}
\includegraphics[width=8.8cm]{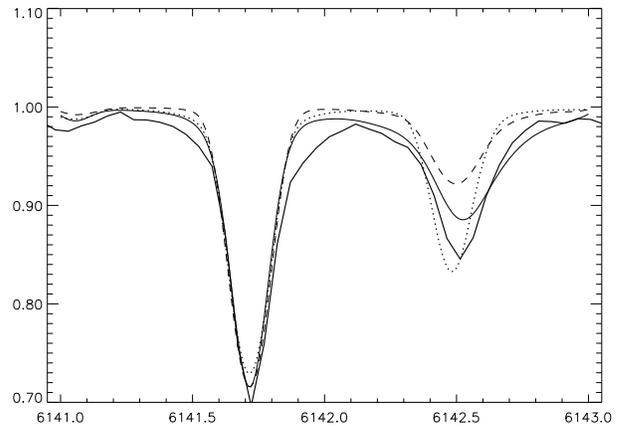}
\caption{The same as in Fig. 5 but for Si\i\, 6142 \AA\ line.}
\end{figure}

\begin{figure}
\includegraphics[width=8.8cm]{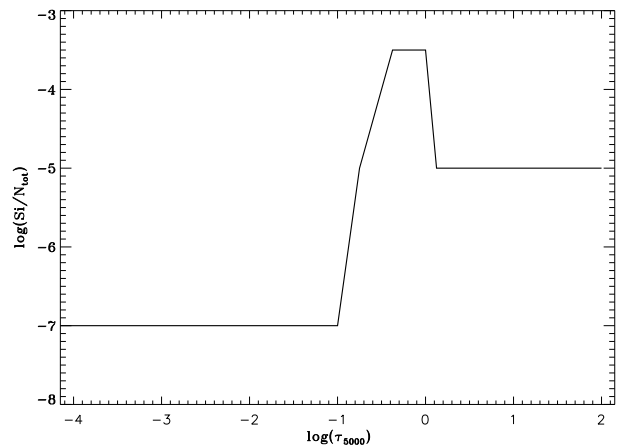}
\caption{Si abundance distribution in the atmosphere of Ap star 10 Aql.}
\end{figure}

\begin{figure}
\includegraphics[width=8.8cm]{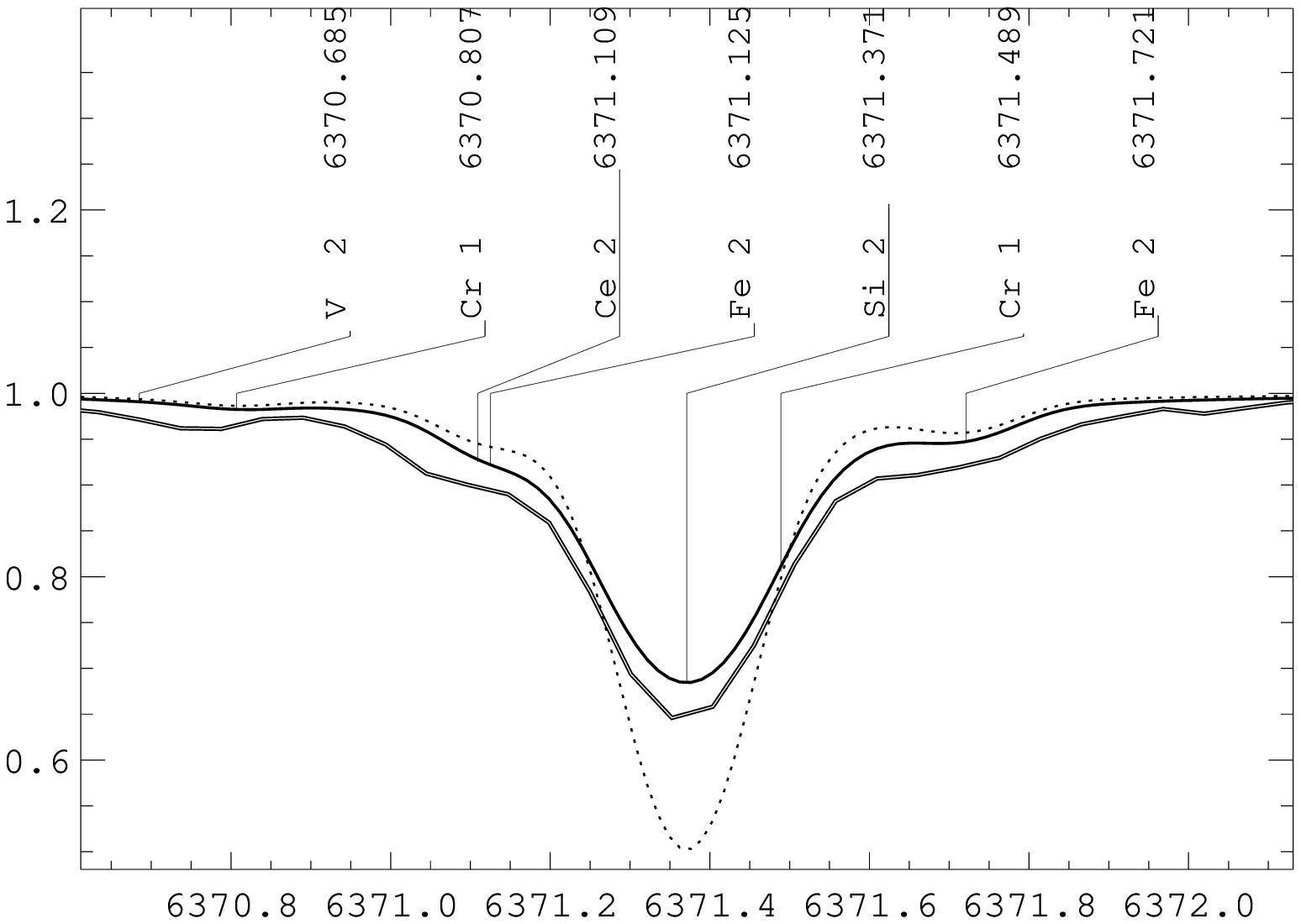}
\caption{A comparison between the observed Si\ii\, 6371 \AA\ line profile
in the spectrum of Ap star 10 Aql (double line) and the synthetic spectrum calculated with 
Si abundance distrubution shown in Fig. 7 (solid line) and Si homogeneous 
abundance
$\log({\rm Si}/N_{tot})$ = -4.19 (dotted line). X- and Y-coordinates are 
wavelenghths and surface fluxes (normalized to unity),
respectivelly}
\end{figure}

In order to simplify the use of Stark broadening data in the codes for stellar
spectral synthesis, we have found an analytical expression for Stark
widths and
shifts

\begin{equation}
{W\over n_e}{\rm [\AA]}=c_1\cdot(A+T^B),
\end{equation}

\begin{equation}
{d\over n_e}{\rm [\AA]}=c_2\cdot(A+C\cdot T^B).
\end{equation}
The
constants $c_1,\ c_2,\ A$, $B$ and
$C$ are given in Table 2. We take $T$  as $T/(10000 K)$.
  The fits of  our calculated data to the analytical
functions given above are satisfactory (Fig. 1) and the differences are
smaller than the expected error of our calculations ($\approx\pm$50\%)

\begin{table}
\begin{center}
     \caption[]{The parameters A, B and C of the approximate formulae 
 for Stark widths and shifts.}
\begin{tabular}{|c|c|c|c|}
\hline
Line & 5950.2 \AA\ & 6142.48\ \AA\ & 6155.13 \AA\ \\
\hline
\hline
\multicolumn{4}{|c|}{WIDTH} \\
\hline
\hline
 \multicolumn{4}{|c|}{electrons} \\
\hline
$c_1$ &1E-16 &1E-14 & 1E-14 \\
A & -0.33801648 &-0.87997007&-0.88824105  \\
B & 0.12759724 & 0.01704956&0.01525209  \\
\hline
\multicolumn{4}{|c|}{protons} \\
\hline
$c_1$ &1E-16 &1E-15 &1E-15 \\
A&-0.81663954 &-0.77664655&-0.76856840  \\
B&0.01728187 &0.03913901&0.03758103  \\
\hline
\multicolumn{4}{|c|}{He\ii} \\
\hline
$c_1$ &1E-16 &1E-15 &1E-15 \\
A&-0.83337682  &-0.81987023&-0.81114334  \\
B& 0.01297813 &0.02771659&0.02946282   \\
\hline
\hline
\multicolumn{4}{|c|}{SHIFT} \\
\hline
\hline
 \multicolumn{4}{|c|}{electrons} \\
\hline
$c_2$ &1E-16 &1E-15 &1E-15 \\
 A & 0.38028502& -0.63997507 &-0.69208568  \\
 B & 1.33496618&0.88440353 & 1.18033516 \\
 C & -0.02047206&0.07950743 & 0.04721782 \\
\hline
\multicolumn{4}{|c|}{protons} \\
\hline
$c_2$ &1E-16 &1E-15 &1E-15 \\
 A &-1.32564473 &1.12387002 &0.17032257 \\
 B &0.01917066 &0.02838789  &0.10512289 \\
 C &1.43394434 &-1.31786299 &-0.37318012 \\
\hline
\multicolumn{4}{|c|}{He\ii} \\
\hline
$c_2$ &1E-16 &1E-15 &1E-15 \\
 A &-1.67370868 &0.14785217 &0.14840180 \\
 B &0.01490401 &0.10161093 &0.10439166 \\
 C &1.75773489 &-0.30183339 &-0.31051296 \\
\hline
\end{tabular}
\end{center}
\end{table}

\subsection{The Stark broadening effect on the shape of Si\i\, lines}

The asymmetry  and shift of Si\i\ 6142.48 \AA\,
and (in particular) 6155.13  \AA\ lines were observed, and it was obvious 
that the
asymmetry is higher for the  hotter stars. The most asymmetrical are 
 $\lambda$ 6142.48 \AA\, and 6155.13 \AA, the strongest lines of $3p^3 \ 
^3D^0-5f^3D$ and  $3p^3 \ 
^3D^0-5f^3G$ multiplets, respectively.
This line is slightly shifted and
shows an asymmetry in the red wing even in a rather cool star like the Sun.
It is not surprising if the Stark effect is responsible for the observed
shifts and asymmetries. From Table 1 one can see that neither widths nor shifts
have a strong temperature dependence, but they depend linearly on 
the perturber density. If a line is strong enough so that line core and
line wings are formed at different atmospheric layers where electron density,
differs significantly, then we should observe line asymmetry
due to differnt shifts at different atmospheric layers even with a
homogeneous vertical distribution of the absorbing element in stellar atmosphere
(the case of the observed asymmetries in normal stars). Vertical abundance
stratification makes this effect more pronounced, and it is observed
in cool Ap stars, where an element stratification takes place
(Savanov~et~al. 2001, Ryabchikova~et~al. 2002, Wade~et~al. 2001). 

Table 3 summarizes the model atmosphere parameters and rotational velocities
for the stars of our sample. These data are taken from Ryabchikova et al. (2000)
and Bikmaev et al. (2002). For the Sun corresponding parameters were taken from
Valenti \& Piskunov (1996) except \teff. Si abundances used in the present 
calculations are given in the last column.

\begin{table*}
\begin{center}
     \caption[]{The atmospheric parameters and rotational velocities
 of the investigated stars.}
\begin{tabular}{|lccccc|}
\hline
Star name       & \teff   & \lgg   & \vt  & \vsini & $\log({\rm Si}/N_{tot})$ \\
or HD           &         &         & \kms & \kms   &                     \\
\hline
  Sun           & 5777    & 4.44    & 0.75       & 1.6  & -4.49 \\
 HD 122970      & 6930    & 4.11    & 0.85       & 5.0  & -4.45 \\
 HD 32115       & 7250    & 4.20    & 2.30       & 9.0  & -4.65 \\
 10 Aql         & 7550    & 4.00    & 0.00       & 5.0  & strat \\
\hline
\end{tabular}
\end{center}
\end{table*}
      
In the Sun the most important broadening is due to collisions with neutral hydrogen
and helium. In the hotter stars the importance of the Stark broadening increases. 
P. Barklem (private communication) has provided us with the value of the broadening parameter
due to collisions with atomic hydrogen for the Si\i\ lines of our interest. This value per 
perturbing particle for a temperature 10000 K, $\log\gamma_{neutral}$ =-6.63, is by 0.4 dex higher
than the corresponding value obtained from the Uns\"{o}ld (1955) approximation. According to P. Barklem
his value may
be overestimated by 0.1 dex which is proved by a comparison between the calculated and the observed blue
wing of the solar Si\i\ lines. The final value used in our calculations is $\log\gamma_{neutral}$ =-6.75.
As for the shift due to collisions with the atomic hydrogen,
a corresponding theory is not well developped, and the value of the shift is rather uncertain
(P. Barklem, private communication). The core and the wings of the strongest Si\i\ $\lambda$ 6155 line
are formed in the layers with the temperatures 5000 and 5800 K respectively. In these layers the corresponding
values for electron-, proton- and atomic hydrogen densities are 4.2  10$^{12}$ 
cm$^{-3}$, 1.5  10$^{11}$ cm$^{-3}$, 
3.9 10$^{16}$ cm$^{-3}$ -- line core, 
and 2.5 10$^{13}$ cm$^{-3}$, 9.3 10$^{12}$ cm$^{-3}$, 1.0 10$^{17}$ cm$^{-3}$ -- line wings. The Stark width is 20 times
smaller than $\gamma_{neutral}$ at the depth of line core formation and  
10 times smaller than $\gamma_{neutral}$ at the depth of line wing formation. In the solar atmosphere, the  
Stark shift is 3 m\AA\ at
the line core and about 20 m\AA\ for the line wings, which is not negligible. In the atmosphere of the 
hottest star of our sample, 10 Aql, the Stark width is of the order of $\gamma_{neutral}$ at the 
depth of line core formation and exceeds $\gamma_{neutral}$ by 5 times at the depth of line wing formation.
Stark shifts are 10 and 100 m\AA, respectively. In 10 Aql the line core is formed at the layers with
$T$=6500 K, $N_{\rm e}$=1.4 10$^{13}$ cm$^{-3}$,  $N_{\rm p}$=1.3 10$^{13}$ cm$^{-3}$, 
$N_{\rm HI}$=5.9 10$^{15}$ cm$^{-3}$ while line wings are formed at the layers
with $T$=7500 K, $N_{\rm e}$=1.2 10$^{14}$ cm$^{-3}$,  $N_{\rm p}$=1.2 10$^{13}$ cm$^{-3}$, 
$N_{\rm HI}$=1.1 10$^{16}$ cm$^{-3}$. 

\subsection{The Sun}

We first calculated Si\i\ lines in the solar spectrum to check the 
Stark parameters. It was mentioned by Ryabchikova~et~al. (2002) that
available theoretical oscillator strengths for the multiplets 
 $3p^3 \ ^3D^0-5f^3D$ and  $3p^3 \ ^3D^0-5f^3G$ are overestimated by about 
0.5 dex
compared to the multiplet $4s^1P^0-5p^1D$ ($\lambda$ 5948.55 \AA). The authors
 fitted the lines of $3p^3 \ ^3D^0-5f^3D$ and  $3p^3 \ ^3D^0-5f^3G$ 
multiplets using the solar
 spectrum. Corrected oscillator strengths -- $\log gf$=-1.42 ($\lambda$ 
6142.48\AA) and
$\log gf$=-0.77 ($\lambda$ 6155.13\AA) were published in Bikmaev~et~al. 
(2002). The latter value was reduced to $\log gf$=-0.82 to get a better fit for
the solar Si\i\ 6155 line.
We also reduced the oscillator strength ($\log gf$=-2.25) of very weak Si line
of the same multiplet, $\lambda$ 6155.69\AA.  
We need to decrease both Stark widths and shifts for
Si\i\ 6142 and 6155 \AA\, lines by 40\% to get a reasonable fit of
the synthetic line profiles to the solar ones. Due to relative weakness of
the Si\i\ 6142 \AA\ line Stark effect is negligible but it is noticeable for the stronger
line Si\i\ 6155 \AA. Fig. 2 shows a comparison between the observed 
profile
of the Si\i\ 6155 \AA\- line in the solar spectrum and synthetic profiles
calculated with (solid line ) and without (dashed line) taking into account 
the Stark width and in particular, shift. The strongest Si\i\ 5948.55 \AA\, line
in our study is fully symmetric as expected taking into account an order of
magnitude smaller values for both the Stark width and shift (see Table 1). 

With corrected Stark parameters we calculated Si\i\ line profiles in spectra
of other stars.  

\subsection{HD~32115}

Spectra of this normal late A-type star have the smallest spectral resolution
of our sample. As it may be seen from Fig. 3 the fit of the 
synthetic
line profile of $\lambda$ 6155.13 \AA\- line to the observed one is much better when
it was calculated with both Stark width and shift. As in the solar case, the Stark
effect is negligible for the weaker Si\i\ 6142.48 \AA\ line. To get a better 
fit of the
line profiles we decrease the Si abundance by 0.13 dex compared to the results
published by Bikmaev~et~al. (2002) which were based mainly on the equivalent width 
measurements. 
 
\subsection{HD~122970}

This is one of the coolest Ap stars and is expected to have a chemically stratified
atmosphere. Results of the abundance determinations show that except for the rare-earth
elements (REE) abundance stratification is marginal if it exists at all (Ryabchikova~et~al. 2000).
Spectral synthesis of Si\i\, lines seems to support this conclusion. We
could reproduce the asymmetric and shifted line profile of Si\i\, 6155.13 \AA\,
reasonably well, using the uniform distribution of Si and Stark 
broadening parameters presented here. Fig. 4 shows a comparison
between the observed and computed line profiles for two cases similar to
Fig. 3. The best fit of Si\i\ line profiles was obtained with the 
same
Si abundance as was derived by Ryabchikova~et~al. (2000).

\subsection{10 Aql}

10 Aql= HD~176232 is the hottest star in our sample. It has the most asymmetrical
Si\i\ 6155.13 \AA\- line profile, which could not be reproduced by any combination
of Stark parameters in a homogeneous atmosphere (Fig. 5). The even weaker line, 
Si\i\
6142.48 \AA\,  shows a noticeable line shift (Fig. 6). 
Ryabchikova~et~al. (2000) mentioned a possibility of Fe and the REE 
stratification in 10 Aql. Therefore we tried to find a simple distribution of Si
in the atmosphere of 10 Aql by trial-and-error which would fit both
Si\i\ 6142.48 \AA\- and 6155.13 \AA\-  lines. The simplest distribution 
shown in Fig. 7
gives a reasonable fit to the observed profiles of both Si\i\ lines 
(Figs. 5 and 6).
Moreover, the same Si distribution seems to fit much better the profiles
of the strong Si\ii\ 6347, 6371 \AA\-  spectral lines compared to the calculations with
the homogeneous Si abundance (-4.19) obtained by Ryabchikova~et~al. (2000).
It is beyond the scope of the present paper to derive an exact Si abundance
profile in the atmosphere of 10 Aql. We  would like to stress here that with 
the present Stark parameters a sensitivity of  6155.13 \AA\ line asymmetry  
to Si abundance changes in the stellar atmosphere can be successfully used in 
empirical studies of abundance stratification in the atmospheres of cool Ap 
stars. 

\section{Conclusions}

In order to discuss the contribution of the Stark broadening effect to the
asymmetry and  the shift of  Si\i\  6142.48\ \AA\ and 6155.13\ \AA\ lines we
have
calculated Stark broadening parameters for these lines by using the
semiclasical perturbation method. The obtained results are the first calculated 
data for the considered
lines. In order to include Stark broadening data we have changed the
STARSP code and we have synthesized the considered
lines. From our analyzis we can conclude that:

1)  The Stark broadening effect is very important for these two lines. The
contribution of electron impact is dominant but, impacts with protons
and He\ii\ ions should be taken into accout as well.

2) The asymmetry as well as the shift of the Si\i\ 6142.48\ \AA\  and 6155.13\
\AA\ lines in many stars including the Sun can be explained by
the Stark broadening effect. 

3) In hotter Ap stars, besides the Stark broadening effect, the stratification
plays a very important role in producing line asymmetry. The sensitivity of the
line asymmetry to changes in the number of Si atoms through the stellar
atmosphere can be used in abundance distribution studies.

\begin{acknowledgements}
We are very grateful to Paul Barklem who provided us with the unpublished
calculations of damping constants for Si\i\ lines due to collisions with 
the atomic hydrogen. 
This work is a part of the projects ``Influence of collisional processes on 
astrophysical plasma lineshapes'' and 
``Astrophysical Spectroscopy of Extragalactic Objects'' supported by the
Ministry of Science, Technologies and Development
of Serbia. The research was supported also by the Fonds zur F\"orderung
der
wissenschaftlichen Forschung {\it P14984} and O\"sterreichische Nationalbank 
({\it Jubila\"aumsfonds Nr.\,7650}).
TR also thanks Russian Federal program ``Astronomy'' and the RFBR (grant
00-15-96722) for partial funding. 
\end{acknowledgements}

\end{document}